\documentclass[aps,twocolumn,showpacs,preprintnumbers,amsmath,amssymb]{revtex4}

\usepackage{graphicx}
\usepackage{dcolumn}
\usepackage{bm}
\usepackage{epsfig}

\begin{document}

\preprint{Draft/\today}

\title{Transport and Thermodynamic Evidence for a Marginal Fermi Liquid State in ZrZn$_2$}

\author{Mike Sutherland$^{1}$, R.P. Smith$^{1}$, N. Marcano$^{1,2}$, Y. Zou$^{1}$, F. M. Grosche$^{1}$, N. Kimura$^{3}$,
S. M. Hayden$^{4}$, S. Takashima$^{5}$, M. Nohara$^{5}$ and H. Takagi$^{5}$}

\affiliation{$^1$Cavendish Laboratory, University of Cambridge, Cambridge, CB3 0HE, United Kingdom,\\
$^2$Centro Universitario de la Defensa, Crta. De Huesca, 50090 Zaragoza, Spain,\\
$^3$Center for Low Temperature Science, Tohoku University, Sendai, Miyagi 980-8578, Japan,\\
$^4$H. H. Wills Physics Laboratory, University of Bristol, Tyndall Avenue, Bristol, BS8 1TL, United Kingdom,\\
$^5$Department of Advanced Materials Science, University of Tokyo, Kashiwa, Chiba 277-8581, Japan}

\date{\today}

\begin{abstract}
Measurements of low temperature transport and thermodynamic properties have been used to 
characterize the non-Fermi liquid state of the itinerant ferromagnet ZrZn$_2$. We observe a $T^{5/3}$ temperature dependence of the electrical resistivity at zero field, which becomes $T^2$ like in an applied field of 9 T. In zero field we also measured the thermal conductivity, and we see a novel linear in $T$ dependence of the difference between the thermal and electrical resistivities. Heat capacity measurements, also at zero field, reveal an upturn in the electronic contribution at low temperatures when the phonon term is subtracted. Taken together, we argue that these properties are consistent with a marginal Fermi liquid state which is predicted by a mean-field model of enhanced spin fluctuations on the border of ferromagnetism in three dimensions. We compare our data to quantitative predictions and establish this model as a compelling theoretical framework for understanding ZrZn$_2$.

\end{abstract}

\maketitle

\section{\label{intro}Introduction}

The Fermi liquid theory of the metallic state is among the most successful in physics,
and decades of research have established its applicability to a wide range of systems.
The fundamental starting point of this theory is the existence of long-lived fermionic quasiparticles whose effective interactions lead only to non diffractive scattering in the zero temperature limit and near to the Fermi surface. In recent years, however, there has been a proliferation of materials that display behaviour not easily understood within the Fermi liquid picture. These so called `non-Fermi liquids' (NFLs) encompass a variety of systems. Examples include the normal state of the high-T$_c$ cuprates \cite{Timusk99}, one-dimensional Tomonaga-Luttinger liquids \cite{Tomonaga50,Luttinger63}, paramagnetic metals near a first order phase transitions \cite{Doiron-Leyraud03} and Kondo lattice systems \cite{Miranda97}. 

A particularly rich variety of NFL behaviour is exhibited by materials in which a magnetic phase transition is observed at low temperatures (see for instance reference \cite{Stewart01}). Tuning the transition to $T \rightarrow$~0~K (a quantum critical point) yields temperature dependences of physical properties that differ from the power laws predicted by Fermi liquid theory. Early attempts to explain this NFL behaviour were based on the effects of strongly enhanced low-frequency, long-wavelength spin fluctuations \cite{Moriya85,Lonzarich97,Hertz76,Millis93,Lonzarich85} (often called the self-consistent renormalisation or SCR theory). In the specific case of three-dimensional materials near a ferromagnetic quantum critical point these approaches suggest characteristic transport and thermodynamic properties. For instance the electrical resistivity is anticipated to vary as $T^{5/3}$ \cite{Mathon68}, instead of the usual $T^2$ behaviour expected in a Fermi liquid, and the electronic heat capacity diverges logarithmically at low temperatures, instead of the Fermi liquid linear in temperature variation.

These non-Fermi liquid temperature dependences are a consequence of an underlying quasiparticle scattering rate, $\tau^{-1}$, which varies linearly with the excitation energy $E$ of a quasiparticle near the Fermi level. This is characteristic not of a Fermi liquid, for which $\tau^{-1}$ varies as $E^2$, but of a marginal Fermi liquid. The study of the marginal Fermi liquid (MFL) state is compelling, as it represents the weakest breakdown of the quasiparticle picture, and thus could prove to be a gateway to understanding more exotic departures from Fermi liquid theory.
  
In this paper we examine in more detail the transport of heat and charge in ZrZn$_2$ reported in a previous work \cite{Smith08}, which is a metal close to a 3D ferromagnetic critical point. We calculate the temperature dependencies of transport properties using SCR theory, and show them to be both qualitatively and semi-quantitatively in agreement with our data. In particular we pay close attention to the difference between thermal and electrical resistivities, following the analytical framework of Paglione $et$ $al.$ \cite{Paglione05} for clean magnetic metals. We then compare these results with new measurements of the electronic specific heat, which we argue lends further support to the validity of the SCR model in this material.  

\section{Experimental Details}

The intermetallic compound ZrZn$_2$ crystallizes in the C15 cubic Laves structure, and is the archetypal d-band itinerant electron ferromagnet \cite{Wohlfarth68}. Improved sample quality as well as reports of the possible co-existence of ferromagnetism and superconductivity \cite{Pfleiderer01,Yelland05b} have stimulated renewed interest in this compound in recent years. The highest quality samples of ZrZn$_2$ order ferromagnetically at $T_c$ = 28.5~K, and magnetization measurements reveal a small moment of 0.17~$\mu_B$/Zr atom in the low-temperature, low-magnetic field limit which is well below the Curie-Weiss moment $\mu_{\mathrm{eff}}$ = 1.9  $\mu_B$/Zr atom \cite{Uhlarz02}. 

The ferromagnetic transition is suppressed to zero with the application of only 20 kbar of hydrostatic pressure \cite{Uhlarz02,Smith08}, suggesting that ZrZn$_2$ at ambient pressure is close to a ferromagnetic quantum critical point. This idea is supported by de Haas-van Alphen studies which show significantly enhanced cyclotron effective masses, as expected in the critical regime \cite{Yates03}, and that longitudinal fluctuations of the local magnetisation are important\cite{Yelland07}. Inelastic neutron scattering studies support the applicability of an SCR model, since the generalised magnetic susceptibility ($\chi_{q,\omega}$) at low wavevector, $q$, and low frequency, $\omega$, has structure characteristic of overdamped diffusive modes with a strongly dispersive relaxation rate \cite{Lonzarich89}.\ 

In our experiments we measured four single crystal samples with varying levels of impurities \cite{Takashima07,Kimura04}. The samples were thin and platelet shaped with typical dimensions of 2 mm $\times$ 0.5 mm $\times$ 0.2 mm. We label each crystal by its residual resistivity $\rho_0$ and residual resistivity ratio (RRR) defined as $\rho_{\mathrm{300K}}$/$\rho_0$. These were measured to be $\rho_0$ = 0.31~$\mu\Omega$cm (RRR = 210), $\rho_0$ = 0.97~$\mu\Omega$cm (RRR = 67), $\rho_0$ = 2.3~$\mu\Omega$cm (RRR = 29) and $\rho_0$ = 6.4~$\mu\Omega$cm (RRR = 11) and were selected to cover a range of impurity concentrations. 

Each sample was spark cut into a convenient geometry, then electropolished to remove a surface layer of approximately 5 $\mu$m in depth. This step was taken to avoid surface inclusions of superconducting material, shown previously to arise from the spark cutting procedure \cite{Yelland05b}. Contacts were made in a four wire geometry using a low power micro-spot-welding technique \cite{Walker98}, yielding contact resistances that were measured to be 5 m$\Omega$ or less at low temperatures. 

Resistivity measurements were performed using a standard 4-terminal low-frequency AC technique in a $^3$He cryostat and in an adiabatic demagnetisation refrigerator, with low-noise transformers to enhance the signal-to-noise level. Measurements were performed on the samples with several different excitation currents between 0.1 and 1mA and several different heating rates to ensure reproducibility.

Thermal conductivity was measured with a two-thermometer, one-heater, steady-state technique down to $T$=0.8~K on a $^3$He system. The reliability of our setup was tested by measuring both heat and charge transport in silver wire, obtaining the Wiedemann-Franz law at low temperatures to within 2\%. Measurements were checked using different models of Cernox thermometers and different sizes of temperature gradient in order to ensure reliability.

The heat capacity measurements between 0.3 and 100~K were conducted in a commercial Quantum Design micro calorimeter at the University of Cantabria using a standard relaxation technique. Only the purest sample was measured ($\rho_0$ = 0.31~$\mu\Omega$cm); a thin slab-shaped crystal weighing 3 mg.

\section{Experimental Results}

\subsection{\label{trandata}Transport Data}

\begin{figure} \centering
              \resizebox{8cm}{!}{
              \includegraphics[angle=0]{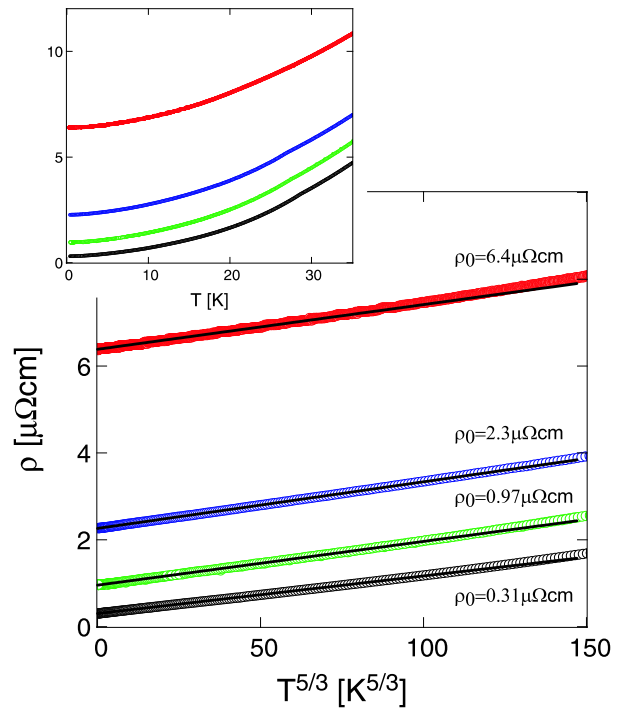}}
             \caption{\label{res}(color online) Main: Resistivity for the four samples of ZrZn$_2$ with plotted versus $T^{5/3}$. The straight line is a fit to the form $\rho$(T)=$\rho_0$ + A$T^{5/3}$ between 0 and 15~K. The inset shows the resistivity plotted against $T$ for a wider temperature range.}
\end{figure}

Figure \ref{res} shows the electrical resistivity of our samples plotted versus $T^{5/3}$. We observe a non-Fermi liquid temperature dependence of the form $\rho(T) = \rho_0+AT^{5/3}$ between 300 mK and approximately 15 K (or higher in the cleanest samples), in agreement with that observed in recent work \cite{Yelland05a, Takashima07}. No evidence of superconductivity was observed down to 300mK in any of our crystals, presumably due to our chemical treatment of the surface. 

The excellent fit to a pure $T^{5/3}$ power law indicates that the scattering of electrons by spin fluctuations is much larger than the scattering by phonons across a wide temperature range. Significant phonon scattering would normally be expected to produce a characteristic $T^5$ temperature dependence in the resistivity at low temperatures, crossing over to a $T$-linear dependence at higher temperatures. Neutron scattering measurements on ZrZn$_2$ are consistent with this observation - the observed scattering intensities are up to two orders of magnitude larger than what could be expected from phonons alone at low temperatures \cite{Lonzarich89}. A similar conclusion was reached in transport studies of the cubic Laves-phase compounds RCo$_2$ (R=Sc,Y,Lu) by Gratz and coworkers \cite{Gratz97,Gratz01}. In these compounds, which lie near a ferromagnetic instability and play host to spin fluctuations, it was concluded that the contribution to $\rho$ from phonon scattering was negligible below 30~K.   

\begin{figure} \centering
              \resizebox{8cm}{!}{
              \includegraphics[angle=0]{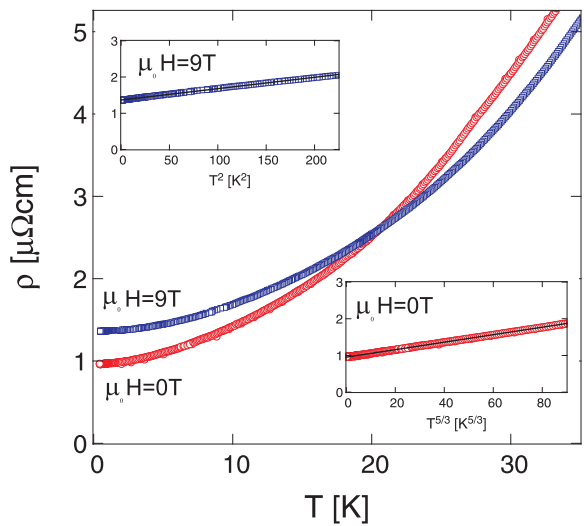}}
             \caption{\label{res_field} (color online) Electrical resistivity of ZrZn$_2$ in a magnetic field. Main panel: $\rho$ vs. $T$ for the sample with $\rho_0$=0.97~$\mu\Omega$cm in $\mu_0H$~=~0 and $\mu_0H$~=~9~T. Upper left inset: $\mu_0H$~=9~T data, plotted vs. $T^2$. Bottom right inset: $\mu_0H$ = 0~T data, plotted vs. $T^{5/3}$. The solid lines in both insets are fits to the data below 15~K.}
\end{figure}

Figure \ref{res_field} shows how the $T^{5/3}$ state is affected by the application of a magnetic field. With $\mu_0H$~=~9~T, the resistivity recovers a characteristic Fermi liquid-like $T^2$ dependence, which may be understood to arise due to the suppression or gapping of the fluctuations in a magnetic field, as suggested by measurements of the electronic specific heat \cite{Pfleiderer01b}. The overall change in the resistivity with temperature is similarly suppressed with field, supporting this interpretation.

In can also be seen from Figure \ref{res_field} that the magnetoresistance of the sample is strongly temperature dependent, as observed previously in polycrystalline samples \cite{Ogawa77}. At low temperatures, the effect is positive, as expected in an ordinary metal. Above $T \sim$ 22~K however the magnetoresistance becomes negative, suggesting that the decrease in resistivity due to the stiffening of spin fluctuations outweighs the increase in resistivity due to orbital magnetoresistance. The recovery of $T^{2}$ power law at high fields helps rule out disorder scattering as the origin of the $T^{5/3}$ power law in $\mu_0H$~=~0, since this process would have little field dependence.

\begin{figure} \centering
              \resizebox{8cm}{!}{
              \includegraphics[angle=0]{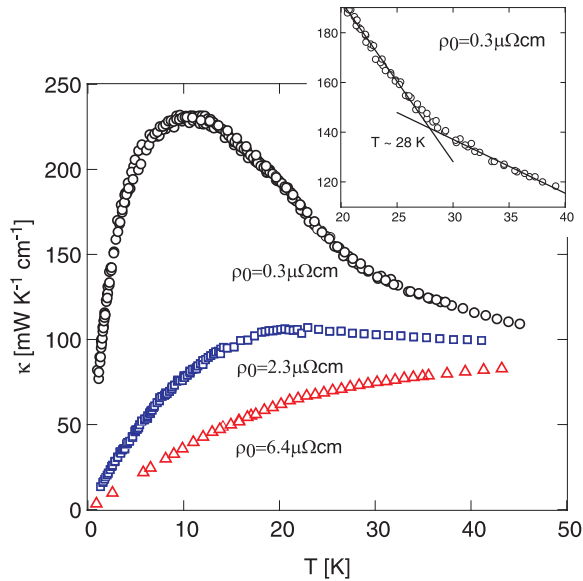}}
             \caption{\label{kappa} (color online) Main: Thermal conductivity of ZrZn$_2$, with samples labeled by their residual resistivities. Inset: Zoom of the transition region in the cleanest sample, with $\rho_0$ = 0.31~$\mu \Omega$cm. Linear fits to the data above and below $T_C$ suggest a change in slope occurs at $T\sim$28~K.}
\end{figure}

Figure \ref{kappa} shows the thermal conductivity of the three ZrZn$_2$ samples. In a metallic material such as ZrZn$_2$, the thermal conductivity has contributions from both electrons ($\kappa_{el}$) and phonons ($\kappa_{ph}$). In the ferromagnetic state at very low temperatures $T<<T_C$, we might also expect a contribution from magnons ($\kappa_{\mathrm{mag}}$). A careful consideration of the relative magnitudes of each of these is necessary to evaluate the suitability of the MFL model in describing ZrZn$_2$.

The thermal current due to coherent, propagating, magnetic modes (magnons) can be difficult to distinguish from other contributions to $\kappa$. Such modes are not the overdamped, diffusive spin excitations mentioned previously, but coherent spin waves existing at very low $q$ that may become populated in the low temperature regime. The contribution to thermal transport from such modes can be considerable, as was demonstrated in the antiferromagnetic insulator Nd$_2$CuO$_4$ \cite{Li05,Zhao11}. However, in magnetic metals $\kappa_{\mathrm{mag}}$ is limited by scattering from charge carriers \cite{Yelon} which can greatly limit the conductivity. We would expect $\kappa_{\mathrm{mag}}$ to be particularly small in ZrZn$_2$ due to its very small ordered moment which limits the phase space available for propagating magnons to $q<q_{sw}$ ($q_{sw}=k_{\uparrow}-k_{\downarrow}\propto M$, where $k_{\uparrow}$ and $k_{\downarrow}$ are the radii of the majority and minority spin sheets of the Fermi surface respectively and $M$ is the magnetisation). We thus ignore the contribution from $\kappa_{\mathrm{mag}}$.

Between the remaining two terms, $\kappa_{el}$ and $\kappa_{ph}$, we argue that by far the largest contribution to thermal conductivity at low temperatures is the electronic term, as might be expected in metals with exceptionally low $\rho_0$. From Figure \ref{kappa}, we see that increasing the residual resistivity from 0.31~$\mu\Omega$cm to 6.4~$\mu\Omega$cm drastically reduces the overall thermal conductivity of the sample, and the characteristic peak seen in the cleanest sample at $T \sim$ 10~K is rapidly suppressed. Since the $\kappa_{el}$ is unlikely to change much between these two samples, we can infer that the reduction in peak height is essentially due to an increase in elastic scattering of electrons from impurities. Studies of controlled levels of impurities doped into conventional ferromagnetic metals have shown a similar trend \cite{Farrell69}.

The inset of Figure \ref{kappa} shows an expanded view of the region near the peak observed in our cleanest sample. A simple linear extrapolation above and below the kink in the data meets at $T \sim$ 28~K, a reasonable estimate for the onset temperature of the peak. This is coincident with the Curie temperature $T_C$ = 28.5~K \cite{Uhlarz02}, and we thus attribute the peak in $\kappa$ to the increase in electronic mean free path as the sample enters the ferromagnetically ordered state and spin fluctuation scattering is reduced.

A more detailed treatment of the relative magnitudes of $\kappa_{el}$ and $\kappa_{ph}$ lends quantitative support to our argument. We can estimate $\kappa_{ph}$ by comparing data from samples with different levels of impurities as follows. We first define the thermal resistivity $w$ using the Wiedemann-Franz law as:

\begin{equation}
\label{define_w}
w = \frac{L_0T}{\kappa_{el}},
\end{equation}
 
\noindent where the Lorentz number, $L_0$ = 1/3($\pi k_B/e)^2 = 2.44 \times$ 10$^{-8}$~W$\Omega$/K$^2$. When considering the transport of both heat and charge by electrons, we must bear in mind that scattering events may affect thermal and electrical currents in different ways.  We thus define the difference between the thermal resistivity ($w$) and the electrical resistivity ($\rho$) as $\delta$, which contains information about scattering processes:

 \begin{equation}
 \delta = w - \rho = \frac{L_0T}{\kappa_{el}} - \rho.
 \label{define_delta}
 \end{equation}

The difference between the effect of scattering on the charge and heat current is that while charge current is only degraded due to the quasiparticle being deflected through a scattering angle $\theta$, the heat current is also degraded by any loss of energy ($\hbar\omega$) of the quasiparticle \cite{Schriempf69,Kaiser71}. The difference between electrical and thermal resistivities can thus reveal information about the nature of inelastic scattering processes, as pointed out by Kaiser for the case of localized magnetic impurities in metallic alloys \cite{Kaiser71} and recently used to investigate antiferromagnetic fluctuations in the clean magnetic metal CeRhIn$_5$ \cite{Paglione05}. In the case of a ferromagnetic metal such as ZrZn$_2$, such inelastic processes are attributed principally to scattering from spin fluctuations. 
 
We may now arrive at an estimate for the phonon conductivity $\kappa_{ph}$. At low temperatures, in keeping with Matthiessen's rule, we assume that the amount of impurities in a given sample in the dilute limit would not considerably alter $\delta$. In other words, we assume that the elastic and inelastic scattering channels are independent.
 
For two samples A and B, with differing impurity levels, we then have:

\begin{equation}
\kappa^A=\frac{L_0 T}{\rho^{A}+\delta}+\kappa_{ph},
\label{sampleA}
\end{equation}
\begin{equation} 
\kappa^B=\frac{L_0 T}{\rho^{B}+\delta}+\kappa_{ph}.
\label{sampleB}
\end{equation}

We assume here that $\kappa_{ph}$ will be the same in each sample at low temperatures, since the presence of impurities is expected to have a minimal effect on low energy phonon modes \cite{Ziman60}. Since we measure both $\kappa$ and $\rho$ for each sample, at every temperature we can then solve these two simultaneous equations to find both $\delta$ and $\kappa_{ph}$. 

       \begin{figure} \centering
              \resizebox{8cm}{!}{
              \includegraphics[angle=0]{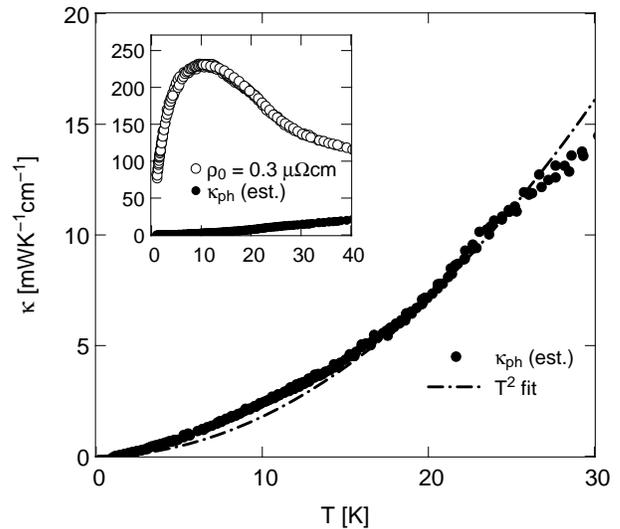}}
             \caption{\label{kappa_ph}The estimated phonon contribution to the total thermal conductivity using the analytical method described in the text. The dashed line represents a fit to $\kappa_{ph}$ $\propto$ $T^2$. The inset shows the estimated phonon contribution compared to the total conductivity measured in the cleanest sample.}
\end{figure}

The solutions of these equations for $\kappa_{ph}$ are shown using the data sets from the samples with $\rho_0$= 0.31 and 6.4~$\mu\Omega$cm in Figure \ref{kappa_ph}. The temperature dependence of $\kappa_{ph}$ below 25~K is reasonably close to the $T^2$ dependence typically expected from phonons scattered by electrons \cite{Ziman60}, and a fit to this form yields a coefficient $\kappa_{ph}/T^2$ $\sim$ 0.018~mWK$^{-3}$cm$^{-1}$. Using different pairs of samples in Equations \ref{sampleA} and \ref{sampleB} yields almost identical curves for $\kappa_{ph}$, differing in magnitude by only 20\%, which demonstrates the consistency of our analysis. 

The phonon contribution that we estimate using this method is seen to be rather small compared to the total conductivity of the least disorder

       \begin{figure} \centering
              \resizebox{8cm}{!}{
              \includegraphics[angle=0]{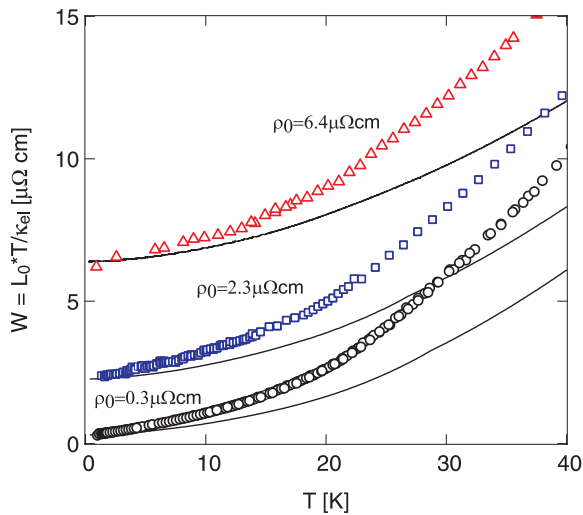}}
             \caption{\label{wandrho}The electrical and thermal resistivity of ZrZn$_2$. The electrical resistivity data are shown as solid lines, while the thermal resistivity data are shown as data points, calculated using $w=$ $L_0 \times T/\kappa_{el}$, where $\kappa_{el}$ = $\kappa_{\mathrm{tot}}$ - $\kappa_{ph}$ as described in the text.}
\end{figure}

Using our estimated form of $\kappa_{ph}$ from Figure \ref{kappa_ph}, we now subtract this from $\kappa_{\mathrm{tot}}$ to find the electronic contribution to thermal transport. For the two samples with the lowest levels of disorder, $\kappa_{ph}$ is small enough so that below $T$=15 K, $\kappa_{\mathrm{tot}}$ $\simeq$ $\kappa_{el}$, meaning $\kappa_{el}$ is insensitive to the details of the phonon subtraction. A similar conclusion was drawn in thermal transport studies below 10 K in the very pure antiferromagnetic metal CeRhIn$_5$ \cite{Paglione05}.

With the above considerations we are in a position to study $\kappa_{el}$ in isolation. In Figure \ref{wandrho} we use $\kappa_{el}$ to calculate the thermal resistivity $w$ defined in Equation \ref{define_w}, which we plot alongside the electrical resistivity $\rho$. All three samples display the same qualitative features: $w \geq \rho$ at finite temperatures, and $w\rightarrow\rho$ in the limit $T \rightarrow 0$ as expected from the Wiedemann-Franz law. 


       \begin{figure} \centering
              \resizebox{8cm}{!}{
              \includegraphics[angle=0]{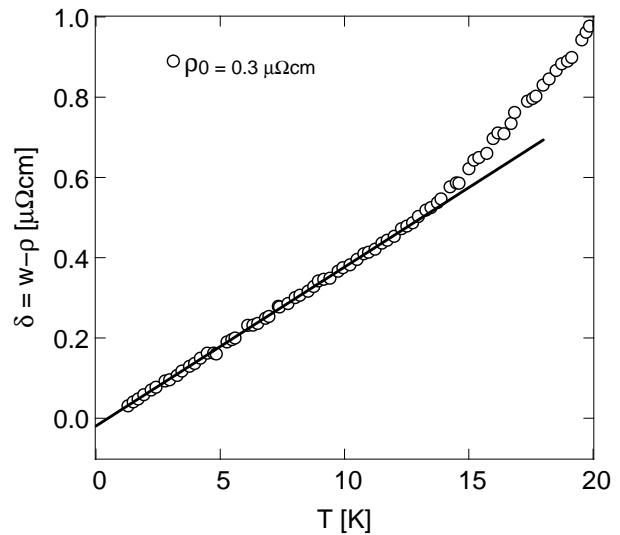}}
             \caption{\label{delta} The difference ($\delta$) of the thermal ($w =L_0T/\kappa_{el}$) and electrical ($\rho$) resistivities  versus temperature for the sample with $\rho_0$=0.31~$\mu\Omega$cm. $\kappa_{el}$ is found by subtracting $\kappa_{ph}$ from $\kappa_{\mathrm{tot}}$, as outlined in the text. The line is a linear fit to the data below 12 K.}
\end{figure}

The difference between the two resistivities is $\delta$, defined in Equation \ref{define_delta} and plotted versus temperature in Figure \ref{delta} for the cleanest sample. The striking linear $T$-dependence of $\delta$ seen in our data persists over an order of magnitude in temperature. In the usual metallic state we would expect $\delta(T)\propto T^2$, since $w = w_0 + AT^2$ and $\rho = \rho_0 + BT^2$. The deviation from this prediction in ZrZn$_2$ is clearly related to the presence of strong inelastic scattering processes operating at low temperatures. We rule out the possibility that $\delta$ is influenced by the presence of phonon scattering at low temperatures since these processes lead to a very different behaviour; in the conventional description phonon scattering leads to a $T^5$ term in $\rho$ and a $T^3$ term in $w$ \cite{Ziman60}. For $T > 12$ K $\delta(T)$ does appear to gain some upward curvature, and we take this temperature to be the point where phonon scattering ceases to be negligible. We note that the same linear behaviour is confirmed in the sample with $\rho_0$=2.3~$\mu\Omega$cm.   

We can compare our results with those obtained in a more conventional magnetic system by Paglione $et al.$, who studied the antiferromagnetic metal CeRhIn$_5$ with T$_N$ = 3.8~K \cite{Paglione05}. In that study, $\delta$ was seen to evolve as $aT^2+bT^5$ below the N\'{e}el temperature and dropped rapidly towards zero above. The $T^5$ term is due to scattering from antiferromagnetic magnons. Clearly the temperature dependence of $\delta$ at low temperatures in ZrZn$_2$ presents an interesting case -- it is difficult to reconcile with either conventional electron-phonon physics or with the model used for CeRhIn$_5$. What we show in Section \ref{Analysis} is that a qualitative and  semi-quantitative understanding of $\delta(T)$ may in fact be gained by considering scattering from spin fluctuations on the border of ferromagnetism in 3D. In other words, the distinct temperature dependence of $\delta(T)$ is a characteristic of the marginal Fermi liquid state.

\subsection{Heat Capacity}

\begin{figure}
\begin{center}
{\includegraphics*[width=8cm]{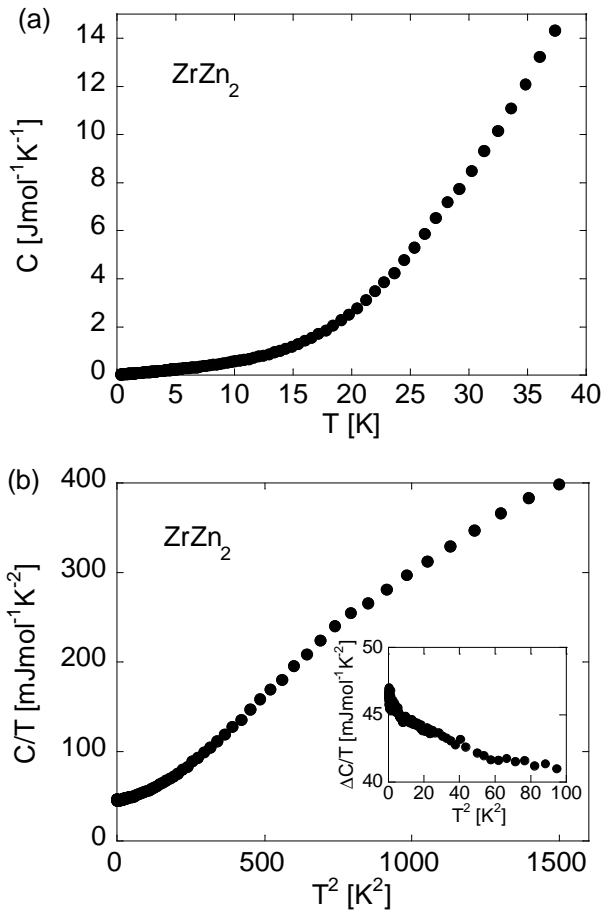}}
\caption{\label{heat}
(a) The heat capacity, $C$, of ZrZn$_2$ versus temperature, $T$. (b) $C/T$ versus $T^2$; the inset shows $\Delta C = (C - C_{ph})/T$ versus $T^2$ where $C_{ph}$ is a $T^3$ Debye-like phonon heat capacity whose slope is given by $\theta_D=340$~K as discussed in the text.}   
\end{center}
\end{figure}

The measured specific heat capacity $C(T)$ of ZrZn$_2$ is shown in Figure \ref{heat}, which are in agreement with those reported previously \cite{Yelland05a}.  Figure \ref{heat}a is a plot of $C$ versus $T$ and Figure \ref{heat}b gives $C/T$ vs $T^2$. The inset of Figure 7b gives $(\Delta C = C - C_{ph})/T$ versus $T^2$, where $C_{ph} = \beta T^3 = (12\pi^4/5)Nk_B(T/\theta_D)^3$ is the phonon contribution to the specific heat at low temperatures in which $\theta_D$ is the appropriate Debye temperature.  In the plot we have taken  $\theta_D$ to be 340~K, a value consistent with previous estimates \cite{Pfleiderer01b}.  We show in the next section that the upturn of  $\Delta C/T$ versus $T^2$ at low temperature is consistent, within experimental error, with the observed temperature dependences of $\delta$  and $\rho$ and with the predictions of the SCR model with parameters relevant to ZrZn$_2$.


\section{\label{Analysis}Discussion}

In this section we consider the possible description of the transport and specific heat data in terms of the SCR model. We begin with an outline of the model and then present the predictions of the model for the model parameters that we believe are relevant to ZrZn$_2$. Finally, the results of the calculations are compared with our transport and specific heat data.

\subsection{Outline of the Self-Consistent Renormalization Model}
\label{SCR}
The theoretical treatment of the effects of the electron-electron interactions on the transport and thermodynamic properties of nearly or weakly ferromagnetic d-metals has a long history (see for instance Refs. \cite{Moriya85,Lonzarich97}).  Here we give only an outline of the key ideas within the self-consistent renormalization (SCR) model mentioned in the introduction.

A discussion of transport properties in this model has been presented most notably by Ueda and Moriya \cite{Ueda75}.  In its simplest form this model involves two isotropic bands.  The carriers in one band, referred to as the $s$-band, carry the current while the electrons in the other band, referred to as $d$-band, scatter the $s$-electrons through a local $s-d$ exchange interaction.  The problem can be formulated in terms of the scattering of $s$-electrons from the spontaneous spin fluctuations arising in the $d$-electron system.  The spin fluctuation spectrum is characterized by a generalized dynamical and wavevector dependent susceptibility which emerges as a central concept in the theory.  The $s$-electrons effectively transfer their momentum to the $d$-electrons.  It is assumed implicitly that the $d$-electrons in turn ultimately transfer their momentum to the lattice via Umklapp processes and scattering from any residual impurities present.

If we ignore residual elastic scattering due to impurities, then for $T<<T_F$ (where $T_F$ is the Fermi temperature for the s-band), $\rho$ and $w$ in this model in the paramagnetic state can be written in the form

\begin{equation}
\label{transport}
\begin{array}{l} \rho \\ w \end{array}\bigg\} =  \frac{\eta}{k_B T} \sum_{q<k_c}\int_0^{\infty}  d\omega \  n_{\omega}(n_{\omega}+1) \frac{\omega}{q} \mathrm{Im} \chi_{q,\omega} \bigg\{ \begin{array}{l} F_{\rho} \\ F_w \end{array}
\end{equation}

where $n_{\omega}$  is the Bose function, $n_{\omega}=(e^{\hbar \omega/k_BT}-1)^{-1}$ and
\begin{equation}
\label{Frho}
F_{\rho}=\left(\frac{q}{k_c}\right)^2
\end{equation}

\begin{equation}
\label{Fw}
F_{w}=\left(\frac{q}{k_c}\right)^2+\left(\frac{\hbar \omega}{k_B T}\right)^2 \frac{3}{4\pi^2}\left(1-\frac{2}{3}\left(\frac{q}{k_c}\right)^2\right) 
\end{equation}
Here $\eta$  is a measure of the s-d coupling parameter, $k_c$ is a wavevector cut-off, equal to the diameter of the $s$-electron Fermi surface in the usual description, and  $\chi_{q,\omega}$  is the generalized wavevector and frequency dependent magnetic susceptibility associated with the $d$-electrons.  The factor $q^2$ (i.e.\ in $F_{\rho}$) arises from the fact that the current cannot be degraded by scattering of $s$-electrons through an infinitesimal angle (i.e.\ $q \rightarrow 0$).  Importantly, the factor $F_w$ for the thermal resistivity includes a second, dynamical term, which allows for the degradation of peak current even when $q$ is vanishingly small through strictly inelastic processes as discussed earlier.  It is this crucial term which leads to the lower temperature exponent in $w$ than in $\rho$. Note that if we set $F$ to 1 then Equation \ref{transport} is proportional to the quasiparticle scattering rate. 

The specific heat in this model arises primarily from the $d$-electrons, and we consider the contribution from dissipative modes \cite{Lonzarich97}. In the paramagnetic state on the border of ferromagnetism at low temperature the specific heat may be expressed within the SCR model approximately in the form
\begin{equation}
\label{cgen}
C=\nu T \frac{d}{dT} \sum_{q<q_c} \int_0^{\infty} d\omega S_{\omega} \mathrm{Im} \left( \frac{\partial \mathrm{log} \chi_{q,\omega}}{\pi \partial \omega} \right),
\end{equation}
where $S_{\omega}$ is the entropy of an undamped boson, i.e. an harmonic oscillator, which is given by
\begin{equation}
S_{\omega}=k_B \left[(1+n_{\omega})\mathrm{log}(1+n_{\omega})-n_{\omega}\mathrm{log}n_{\omega} \right].
\end{equation}
The parameters $\nu$  and $q_c$ correspond to the degeneracy of spin fluctuation modes and the wavevector cut-off respectively.  In the usual description  $\nu = 3$ and $q_c$ is the diameter of the $d$-electron Fermi surface.

To evaluate the expressions for $\rho$, $w$ and $C$ we need in particular Im$\chi_{q,\omega}$, which we model using data from inelastic neutron scattering experiments.  The cut off wavevectors $k_c$ and $q_c$ can be estimated from the known properties of the Fermi surface, measured in quantum oscillation experiments \cite{Yates03}. The parameter $\eta$  can in principle be inferred from the Kadowaki-Woods relationship that has been shown \cite{Moriya95} to be consistent with the SCR model under conditions where this model reduces to the Fermi liquid limit.  Since we shall be mainly interested in the relative magnitudes of $\rho$  and $w$ we shall for simplicity scale both quantities to a reference value of $\rho$ at 15 K (a crossover temperature above which phonon contributions to the transport properties become progressively less ignorable). Although we shall focus on the relative values of $\rho$ and $w$, we note that by applying the above Kadowaki-Woods relationship we obtain in our calculations absolute values of $\rho$ and $w$, which are within 20\% of the measured values for the model parameters relevant to ZrZn$_2$.

Within the SCR model $\chi_{q,\omega}$ in the paramagnetic state may be expressed in the form
\begin{equation}
\chi_{q,\omega }^{-1}=\chi^{-1}(T)+cq^2-i\frac{\omega}{\gamma q},
\label{dynsus}
\end{equation}
where  $\chi(T)$ is the static magnetic susceptibility, and $c$ and $\gamma$ are constants dependent on the details of the $d$-electron band.  The temperature dependence of the static susceptibility can in principle be determined self-consistently within the SCR model.  Here we take $\chi(T)$ from bulk susceptibility data and the parameters $c$ and $\gamma$ from inelastic neutron scattering data, which could be satisfactorily described by the model given by Equation \ref{dynsus} \cite{Lonzarich89}.

In the ferromagnetic state below the Curie temperature $T_C$ we must distinguish between the components of the spin fluctuations parallel ($\|$) to the ordered moment and perpendicular ($\bot$) to the ordered moment.  The principal changes are as follows.  Firstly, the static susceptibility $\chi(T)$ becomes strongly dependent on the component (there are two transverse ($\bot$) components and one longitudinal ($\|$) component).  Secondly, the   component of the generalized susceptibility has a spin-wave form below the cut-off $q_{sw}$ defined earlier.  In this regime we have \cite{Lonzarich97}
\begin{equation}
\label{sw_chi}
\chi_{\bot q,\omega}^{-1}=cq^2(1-\omega^2/\Omega_q^2)
\end{equation}
where
\begin{equation}
\label{sw_disp}
\hbar \Omega_q=Dq^2=2\mu_B Mcq^2
\end{equation} 
is the spin wave spectrum at small $q$.

\begin{figure} \centering
              \resizebox{8cm}{!}{
              \includegraphics[angle=0]{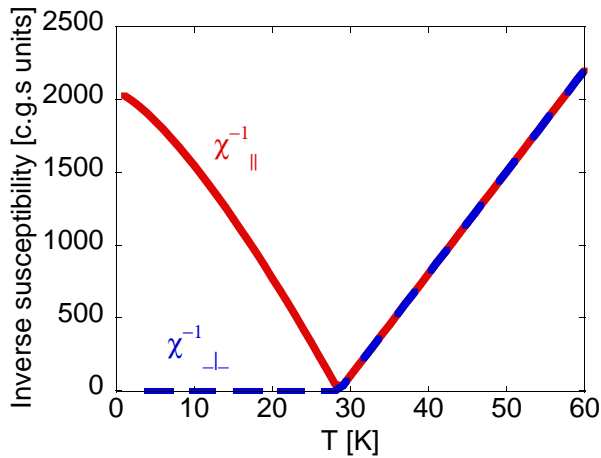}}
             \caption{\label{fig:5}(color online) Plots showing the form of the dimensionless inverse volume susceptibility (in c.g.s units of Oe/(emu cm$^{-3}$)) used in the resistivity calculations for ZrZn$_2$. The expressions of these are as follows: $\chi_{\|}^{-1}=2(36T_C(1-(T/T_C)^{4/3}))$ and $\chi_{\bot}^{-1}=0$ for $T\leq T_C$ and $\chi_{\|}^{-1}=\chi_{\bot}^{-1}=70(T-T_C)$ for $T>T_C$. Here we take $T_C=$28.5~K.}
\end{figure}

From inelastic scattering data for ZrZn$_2$ we have approximately  $\gamma$=1.8~$\mu$eV\AA \ and $c$= 3$\times 10^5$ ~\AA$^2$ \cite{Lonzarich89}.  From the measured Fermi surface of ZrZn$_2$ we estimate $q_{sw}=$ 0.07~\AA$^{-1}$ \cite{Yates03}.  From bulk measurements of the magnetic equation of state we arrive at models for $\chi_{\bot}(T)$ and  $\chi_{\|}(T)$ as shown in Figure \ref{fig:5} and $M(T)= M_0(1 - T^2/T_C^2)^{1/2}$ where $M_0 =$31 G.  We note that the results of the calculations are not sensitive to precise forms of $\chi$ and $M$ or even to the precise values of the other parameters.  For non-Fermi liquid behaviours to arise in this model we require principally that $\chi^{-1}$ is small compared with $cq^2$ for the characteristic wavevectors for thermally excited spin fluctuations.  In other words, we require that the characteristic thermal wavevector, $q_T$, is large compared with the magnetic correlation wavevector (i.e., the inverse of the magnetic correlation length $\xi = \sqrt{\chi/c}$) \cite{Lonzarich97}.

In practice, d-metals such as ZrZn$_2$ cannot be represented simply in terms of two distinct species of electron as the above model suggests and the terms ``$s$-electron" and ``$d$-electron" may not have precise meanings.  It is of interest, however, to see if the model can capture some of the behaviours of $\rho$, $w$ and $C$, which we have measured in ZrZn$_2$.

It has been emphasized in more recent theoretical works that there should be non-analytic corrections to certain aspects of the SCR model \cite{Belitz97, Chitov01, Chubukov06}.  Although relatively weak in 3D, they can nevertheless strongly modify the magnetic equation of state in particular and give rise to a first-order ferromagnetic transition near the quantum critical point and potentially to a magnetically inhomogeneous state.  We shall therefore not attempt to calculate the magnetic equation of state self-consistently in this paper and take the temperature dependence of aspects of the magnetic equation of state from experiment.


\subsection{The Marginal Fermi Liquid Limit of the SCR Model}
\label{logFL}
In the limit $q_T >> \xi^{-1}$ it may easily be shown from Equations \ref{transport} - \ref{dynsus} that in 3D
\begin{equation}
\rho \propto T^{5/3} 
\label{log_rho}
\end{equation}

\begin{equation}
\delta=(w-\rho) \propto T
\label{log_delta}
\end{equation}

\begin{equation}
\tau^{-1}\propto T \mathrm{log}(T^*/T)
\end{equation}
and

\begin{equation}
C \propto T \mathrm{log}(T^*/T)
\end{equation}
where $T^*$ is an appropriate temperature scale in each case separately. Note that the temperature dependence of the scattering rate  $\tau^{-1}$ is different from that of $\delta$ due to the $(\omega/T)^2$ factor in Equation \ref{Fw}.  Thus, in this limit, the SCR model reduces to the MFL model discussed in the introduction, while in the opposite limit, $q_T << \xi^{-1}$, the SCR model reduces to the FL model.  For the parameters relevant to ZrZn$_2$ we find that the crossover to the FL model occurs only at temperatures below 1 K at ambient pressure\endnote{This should be contrasted with the analogous problem in MnSi where in the same model, but with parameters relevant to MnSi, the FL crossover temperature is an order of magnitude higher.  In MnSi, the MFL regime is squeezed out by the FL domain at low temperatures and the phonon contribution at higher temperatures.  In ZrZn$_2$ the MFL is then expected to be a good approximation to the SCR model over a wide temperature range, up to over 50~K (with the exception of a narrow region close to $T_C$) if the phonon contributions can be ignored.  The phonon contribution ignored in the SCR model in practice limits the applicability of the MFL to below about 15~K in ZrZn$_2$.}.

The characteristic marginal Fermi liquid exponents for $\rho$  and $\delta$ can be understood by analogy to the corresponding electron-phonon scattering problem.  In that problem $\rho \propto q_T^{d+2}$ and $\delta \propto q_T^{d}$ , where $d$ is the dimension of space. The extra factor of $q_T^2$ in $\rho$  comes from the $q^2$ term in $F_{\rho}$ (Equation \ref{Frho}) which is absent in the difference between $w$ and $\delta$.  The temperature dependence of $q_T$ is given by the dynamical exponent $z$ via $T \propto q_T^z$ so that one expects
\begin{equation}
\rho \propto T^{\frac{d+2}{z}}, 
\end{equation}
\begin{equation}
\delta \propto T^{\frac{d}{z}}. 
\end{equation}

For the electron-phonon scattering problem $z = 1$ and in 3D we recover the usual results, $\rho \propto T^5$ and $\delta  \propto T^3$.  For electron paramagnon scattering, however, z = 3 (see Equation \ref{dynsus} in the limit $\chi^{-1}\rightarrow 0$) and so in 3D we expect  $\rho \propto T^{5/3}$ and $\delta  \propto T$ as in Equations \ref{log_rho} and \ref{log_delta}.  We caution that for dissipative modes such as paramagnons this elementary treatment fails if $z$ is unity or lower.  In this case,  $\rho$ and $\delta$ are determined by spin-fluctuations over all of $q$-space rather than just at low $q$ and to obtain the FL limit of the SCR model one must return to the full expressions for $\rho$  and $w$ in Equation (\ref{transport}).

\subsection{Calculations of the Electrical and Thermal Resistivities}

\begin{figure} \centering
              \resizebox{8cm}{!}{
              \includegraphics[angle=0]{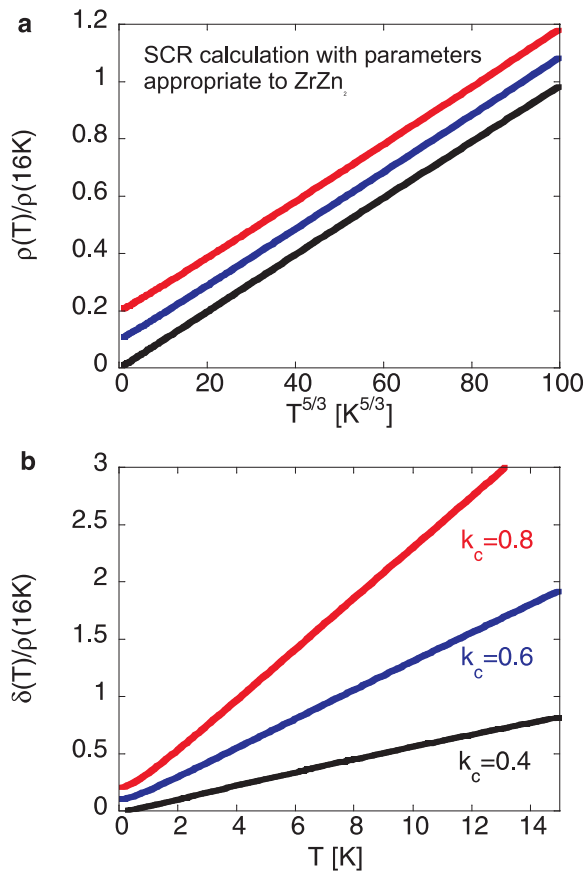}}  
             \caption{\label{transcalc}(color online) Plot showing (a) the calculated electrical resistivity plotted against $T^{5/3}$ and (b) the calculated difference between the thermal and electrical resistivities plotted against $T$. The results are normalised relative to the calculated electrical resistivity at 16 K.  The results are shown for three values of the wavevector cut-off parameter $k_c$ (0.4~\AA$^{-1}$ (bottom), 0.6 \AA$^{-1}$ (middle) and 0.8~\AA$^{-1}$ (top)).  The data for $k_c=0.6$~\AA$^{-1}$ and $k_c=0.8$~\AA$^{-1}$ are shifted upwards by 0.1 and 0.2 respectively, for clarity.}
\end{figure}

Using the model and parameters given above we can calculate the electrical and thermal resistivities and the difference $\delta = w -\rho$.  The results of these calculations are shown in Figure \ref{transcalc} for several values of $k_c$ (which in the model represents the typical diameter of the Fermi surface of the light electrons that principally carry the current).


The value of $k_c$ that gives the correct magnitude of the scaled quantity $\delta(T)/\rho$(15~K) turns out to be around $k_c$=0.4~\AA$^{-1}$.  It is interesting to note that this is close to the observed diameter of the sheet of the Fermi surface of ZrZn$_2$ with the highest characteristic Fermi velocity observed in this system.  However, this fitted value of $k_c$ should not be taken too seriously, given the complexity of the Fermi surface of ZrZn$_2$ and the relative simplicity of the present model.  What seems to be significant, however, is that the calculated temperature dependence of $\rho$  and $\delta$  are qualitatively very similar to the observed values given in Section \ref{trandata}, irrespective of the precise value chosen for $k_c$, or indeed of the precise values of the other model parameters.

Within the experimental temperature range of interest, we find that the SCR model reduces essentially to the MFL model as discussed in the introduction.  In particular, $\rho$  is predicted to vary approximately as $T^{5/3}$ and $\delta$  approximately as $T$ in the temperature range below 15~K (above which phonon contributions become progressively more important in the measurements) and above about 1~K (well below which the SCR model predicts a Fermi liquid form for $\rho$ and $w$).

\subsection{Calculations of the Specific Heat}

\begin{figure} \centering
              \resizebox{8cm}{!}{
              \includegraphics[angle=0]{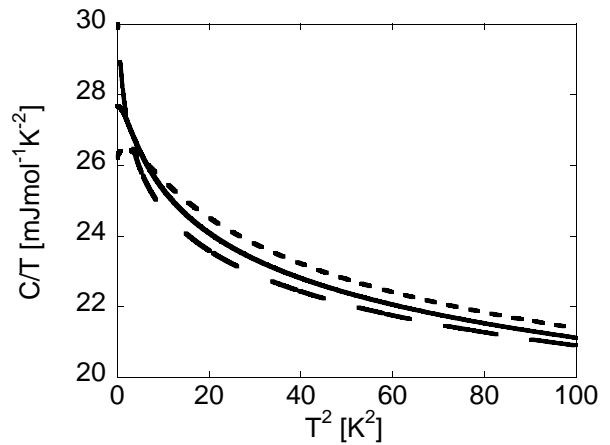}}
             \caption{\label{comp}
Plot showing the calculated spin fluctuation heat capacity for several values of $q_{sw}$, namely 0.05~\AA$^{-1}$, 0.08~\AA$^{-1}$ and 0.1 \AA$^{-1}$. The effect of increasing $q_{sw}$ is to cut off the low temperature divergence of $C/T$.}
\end{figure}

Using the model and parameters in Section \ref{SCR} we can also calculate the specific heat contribution of the spin-fluctuations.  Below $T_C$ we consider the contribution from both dissipative modes and from spin waves with $q<q_{sw}$. The results of the calculations are shown in Figure \ref{comp} in the form of $C/T$ vs $T^2$ for comparisons with the inset of Figure \ref{heat}b of the measured specific heat minus the estimated phonon contribution.  The calculated heat capacity depends approximately logarithmically on the upper cut-off $q_c$ and is also somewhat sensitive to the spin-wave cut-off $q_{sw}$, although the relative contribution from the dissipative modes with $q>q_{sw}$ dominates at all temperatures. The results in Figure \ref{comp} correspond to the choice $q_c$= 1~\AA$^{-1}$, which is approximately equal to the observed diameter of the large sheets of the Fermi surface with high effective mass (and so low Fermi velocity).

To check on the sensitivity of the result to the lower spin wave cut-off several values of $q_{sw}$  around the value of 0.07~\AA$^{-1}$ given in Section \ref{SCR} have been used.


Comparing the inset of Figure. \ref{heat}b with Figure \ref{comp} we see that the measurements are qualitatively consistent with the predictions of the SCR model, especially for a reasonable value for $q_{sw}$.  In particular, the upturn in $C/T$ with decreasing temperature in the inset of Figure 7b is consistent with the predictions of the SCR model.  In the limit $q_T  >> \xi^{-1}$ (see Section \ref{logFL}) this upturn is reminiscent of the logarithmic term in $C$ expected for a marginal Fermi liquid.  The magnitude of the calculated $C/T$ is somewhat lower than the measured value. This is reasonable since the SCR model does not include the effects of the electron-phonon interaction, nor, in the form defined in Section \ref{SCR}, the effects of any antiferromagnetic spin fluctuations that may be present.
\vspace{20pt}
\section{Conclusions}

In summary, we find that the temperature dependences of the electrical and thermal resistivities and the electronic specific heat in ZrZn$_2$ can be understood qualitatively and even semi-quantitatively in terms of the self-consistent renormalization model with model parameters inferred from independent measurements of the magnetic equation of state, the inelastic neutron scattering cross-section and the Fermi surface.  Over a wide range in temperature, the SCR model reduces to the marginal Fermi liquid model characterized by an electrical resistivity $\rho$ varying as $T^{5/3}$, a thermal resistivity $w$ - $\rho$ varying as $T$ and a logarithmic divergence in the low temperature specific heat. We believe that taken together, these findings provide the most compelling evidence thus far for the existence of the marginal Fermi liquid state in the field of itinerant-electron ferromagnetism.

\section{Acknowledgments}

We thank Gil Lonzarich, Montu Saxena, Lara Sibley, Ed Yelland and Stephen Rowley for useful discussions. This work was funded by the Royal Society and EPSRC. The authours wish to thank the following organizations for financial support - R. P. Smith acknowledges St. Catherines college, University of Cambridge, N. Marcano acknowledges financial support from the MAT2008-06542-C04 project, and M. Sutherland acknowledges the Royal Society. Special thanks go to I. R. Walker for his help in setting up the Oxford Instruments Heliox system.

\bibliography{ZrZn2_submitted}

\end{document}